\documentclass[referee, sn-basic]{sn-jnl}%

\usepackage{graphicx}%
\usepackage{multirow}%
\usepackage{amsmath,amssymb,amsfonts}%
\usepackage{amsthm}%
\usepackage{mathrsfs}%
\usepackage[title]{appendix}%
\usepackage{xcolor}%
\usepackage{textcomp}%
\usepackage{manyfoot}%
\usepackage{booktabs}%
\usepackage{algorithm}%
\usepackage{algorithmicx}%
\usepackage{algpseudocode}%
\usepackage{listings}%
\usepackage{array}

\usepackage[normalem]{ulem}

\theoremstyle{thmstyleone}%

\theoremstyle{thmstyletwo}%

\theoremstyle{thmstylethree}%

\begin{document}

\title[Network Analysis of U.S. Non-Fatal Opioid-Involved Overdose Journeys, 2018-2023]{Network Analysis of U.S. Non-Fatal Opioid-Involved Overdose Journeys, 2018-2023}

\author*[1,2]{\fnm{Lucas H.} \sur{McCabe}}\email{lucasmccabe@gwu.edu}

\author*[3,4]{\fnm{Naoki} \sur{Masuda}}\email{naokimas@buffalo.edu}

\author[5]{\fnm{Shannon} \sur{Casillas}}

\author[1]{\fnm{Nathan} \sur{Danneman}}

\author[5]{\fnm{Alen} \sur{Alic}}

\author[5]{\fnm{Royal} \sur{Law}}

\affil*[1]{\orgname{LMI}, \orgaddress{\street{7940 Jones Branch Drive}, \city{Tysons}, \postcode{22102}, \state{VA}, \country{USA}}}

\affil[2]{\orgdiv{Department of Computer Science}, \orgname{The George Washington University}, \orgaddress{\street{800 22nd Street NW}, \city{Washington}, \postcode{20052}, \state{DC}, \country{USA}}}

\affil[3]{\orgdiv{Department of Mathematics}, \orgname{State University of New York at Buffalo}, \orgaddress{\street{244 Mathematics Building}, \city{Buffalo}, \postcode{14260}, \state{NY}, \country{USA}}}

\affil[4]{\orgdiv{Institute for Artificial Intelligence and Data Science}, \orgname{State University of New York at Buffalo}, \orgaddress{\street{215 Lockwood Hall}, \city{Buffalo}, \postcode{14260}, \state{NY}, \country{USA}}}

\affil[5]{\orgdiv{National Center for Injury Prevention and Control}, \orgname{Centers for Disease Control and Prevention}, \orgaddress{\street{4770 Buford Highway, NE}, \city{Atlanta}, \postcode{30341}, \state{GA}, \country{USA}}}

\abstract{We present a nation-wide network analysis of non-fatal opioid-involved overdose journeys in the United States. Leveraging a unique proprietary dataset of Emergency Medical Services incidents, we construct a journey-to-overdose geospatial network capturing nearly half a million opioid-involved overdose events spanning 2018-2023. We analyze the structure and sociological profiles of the nodes, which are counties or their equivalents, characterize the distribution of overdose journey lengths, and investigate changes in the journey network between 2018 and 2023. Our findings include that authority and hub nodes identified by the HITS algorithm tend to be located in urban areas and involved in overdose journeys with particularly long geographical distances.}


\maketitle

\section{Introduction}

The toll of the drug overdose epidemic in the United States is staggering. More than one hundred thousand people lost their life from a drug overdose in 2021 alone \citep{spencer2022drug}. Opioids have been a major driver of the epidemic, and opioid overdose mortality in the U.S. has grown exponentially since $1979$ \citep{jalal2018changing}. Furthermore, the rate of nonfatal drug overdose Emergency Medical Services (EMS) encounters involving opioids nearly doubled from January 2018 to March 2022 \citep{casillas2022patient}. Evidence suggests that state-level efforts have been moderately successful in reducing misuse of prescription opioids \citep{dart2015trends}, but such policies may have unintentionally increased opioid mortality by indirectly incentivizing illicit usage \citep{lee2021systematic}. For every fatal drug overdose there are many more nonfatal overdoses, and EMS data uniquely provide information on where nonfatal overdoses occur, as well as details on patient residence. This information can be used for understanding how far from a person’s place of residence they experienced an overdose, hereafter called a journey to overdose.

Network analysis has been proven to be useful for studying multiple types of public health questions, including disease transmission, diffusion of health-related behavior and information, and social support networks, which are relevant to opioid overdose \citep{luke2007network, Luke2012AnnuRevPublHealth, Valente2017AnnuRevPublHealth}. In the context of the opioid overdose crisis, social network analysis of Appalachians found that over half the individuals in their data set had a first-degree relationship with someone who experienced an opioid overdose, and this proportion was higher near city centers \citep{rudolph2019using}. It was also found that measures of network centrality and prominence can elucidate illicit opioid-seeking behavior \citep{perry2019co, yang2022comparing}.

The journey to overdose stems from the so-called ``journey to crime," a framework for conceptualizing mobility patterns in environmental criminology. 
Typically, each event is associated with an origin (e.g., home), destination (e.g., site of the offence), and intermediate phase, wherein an individual is assumed to have traveled from the origin to the destination \citep{rengert2002journey}, indicating that network science tools may be appropriate for such analyses.
Although journey-to-crime distances are often fairly short \citep{phillips1980characteristics}, recent work has identified offense and offender categories for which journey-to-crime distances tend to be longer \citep{vandeviver2015makes, van2010journey, chen2021journey}, suggesting that long journeys may be worth examining in their own right.

To the best of our knowledge, Oser et al.\,applied the concept of geographical discordance to drug overdoses for the first time~\citep{oser2015treatment}. They found that people who use drugs who traveled to counties other than those of their residences for substance abuse treatment are more likely to relapse with prescription opioids. They also found that geographically discordant treatment efforts (i.e., when those in treatment obtain it in a county other than that of their residence) are more common among rural populations than suburban or urban ones. Johnson et al.\,studied trip distance to drug purchase arrest for several classes of controlled substances, leveraging arrest record data from Camden, New Jersey. They showed that race and ethnicity were significantly correlated with trip distance, underscoring the relationship between geospatial segregation and drug purchase patterns \citep{johnson2013need}.
Donnelly et al. examined trips leading to opioid possession arrest in Delaware, finding dependency of journey length on age, gender, and racial covariates \citep{donnelly2021opioids}.
Forati et al.\,applied spatial network methods to analyze geographically discordant opioid overdoses, i.e., those that occur in locations distinct from the residence of the person who experienced the overdose. Their work, which concerned fatal incidents in Milwaukee, Wisconsin between 2017 and 2020, was the first to analyze an overdose journey \textit{network} \citep{forati2023journey}.
However, the trips quantified in these studies were relatively short, and the study included only a single area. In the present study, we extend these works to a larger spatial scale (i.e., across the U.S.) and to a time horizon that includes the COVID-19 pandemic.

\section{Materials and methods}\label{sec:datameth}

\subsection{Data}

This section introduces data and several domain-specific acronyms that may not be familiar to all audiences. The acronyms are itemized in Table \ref{tab:acronyms} for readers' convenience.

\subsubsection{EMS data}\label{sec:ems}

We utilize a proprietary dataset sourced from biospatial, Inc., a platform specializing in the collection and analysis of EMS data \citep{biospatial}. Company biospatial aggregates information from hundreds of counties across 42 U.S. states, including 27 with full coverage and 15 with partial coverage. ``Full coverage'' states are those from which biospatial receives all records from the state EMS office, while ``partial coverage'' states provide a subset of the data, sometimes garnered through direct partnerships with EMS providers. Place-of-residence information is collected from the patient, patient's history, or from the EMS destination (e.g., hospital). Data from EMS providers are available at the patient level, but exact information on overdose location and patient residence are not available to external researchers due to privacy protections. We obtain from biospatial a version of the dataset that has been spatially aggregated to protect patient identities. We have visibility of overdose incident locations and patient residences in terms of their encompassing counties or county equivalents (for simplicity, we also refer to them as counties in the following text). We represent counties by their corresponding Federal Information Processing Standards (FIPS) codes, which are five-digit numeric codes assigned by the National Institute of Standards and Technology (NIST) to uniquely represent counties within the U.S.

The Council of State and Territorial Epidemiologists (CSTE) is a U.S. non-profit public health organization providing the Nonfatal Opioid Overdose Standard Guidance, a procedure for identifying suspected opioid-involved overdose encounters by querying coded data elements and patient care report narratives. Specifically, it considers provider impressions, symptoms observed, medications administered and the subsequent response, and keywords considered opioid-related or overdose-related. 
Following this protocol, we sub-set our data to nonfatal opioid occurrences and remove request cancellations and EMS call-outs coded as assisting another primary agency \citep{cste2022ems}.

Coverage and completeness of biospatial's EMS data vary by time period and reporting agency. Company biospatial maintains an internal probabilistic model, based on population characteristics and historical data, that calculates an Underlying Event Coverage (UEC) metric. UEC estimates the portion of total EMS events during a given time period in a county that are captured by the biospatial platform.
Therefore, our first exclusion criterion is counties without sufficient data. Following Casillas et al., we only include counties with a UEC of at least 75\% during each quarter of the study period, in part to mitigate bias associated with under-reporting \citep{casillas2022patient}. 
We briefly assess robustness of the results of the downstream analysis with respect to this threshold in Section \ref{sec:sensitivity}.
Our second exclusion criteria includes counties that do not share Patient Care Report Narratives with biospatial, limiting the ability of the CSTE opioid overdose definition queries to identify encounters. Lastly, we excluded fatal opioid occurrences and focused on non-fatal opioid-involved overdose encounters.
The geographical distribution of event coverage in the dataset is shown in Figure \ref{fig:uec_map}.

\subsubsection{Other data}\label{other_data}

We obtain demographic and socioeconomic statistics from the American Community Survey 2017-2021 5-Year Data Release (ACS). The ACS is an annual demographic survey managed by the U.S. Census Bureau, which provides details on education, employment, housing, and other demographic factors. The survey samples approximately 3.5 million addresses from across the U.S. annually \citep{acs2021}.

We also use urbanicity data from the 2013 National Center for Health Statistics (NCHS) Urban–Rural Classification Scheme for Counties (URCS), which categorizes U.S. counties into the following six groups based on urbanization:
\begin{itemize}
    \item \textbf{Large central metro}: counties that (i) are in a metropolitan statistical area (MSA) with at least one million residents, (ii) contain the entirety of the MSA's largest principal city, and (iii) are exclusively populated within the MSA's largest principal city or represent at least 250000 inhabitants in one of the MSA's principal cities,

    \item \textbf{Large fringe metro}: counties in an MSA that have at least one million residents and are not classified as large central metro. These counties are, for example, the suburbs of large central metros,
    
    \item \textbf{Medium metro}: counties in an MSA with between $250000$ and $999999$ residents,
    
    \item \textbf{Small metro}: counties in an MSA with fewer than $250000$ residents,
    
    \item \textbf{Micropolitan}: counties in micropolitan statistical areas, and
    
    \item \textbf{Noncore}: counties in neither MSAs nor micropolitan statistical areas \citep{ingram20142013}.
\end{itemize}
As in \citep{casillas2022patient}, we refer to large central metropolitan, large fringe metropolitan, medium metropolitan, or small metropolitan counties as being urban, and micropolitan and noncore counties as being rural, for simplicity. We also provide results for all six categories in Appendix~\ref{sub:6-category-results}.

The U.S. Census maintains the Topologically Integrated Geographic Encoding and Referencing (TIGER) system \citep{weather_gov_counties}, which provides cartographic information about U.S. geographic areas. We extract county boundary coordinates from TIGER's shape files to support our analysis of overdose journey distances.

\subsection{Construction of a nationwide spatial network of overdose journeys}

\begin{figure}[htbp]
    \centering
    \includegraphics[width=1\textwidth]{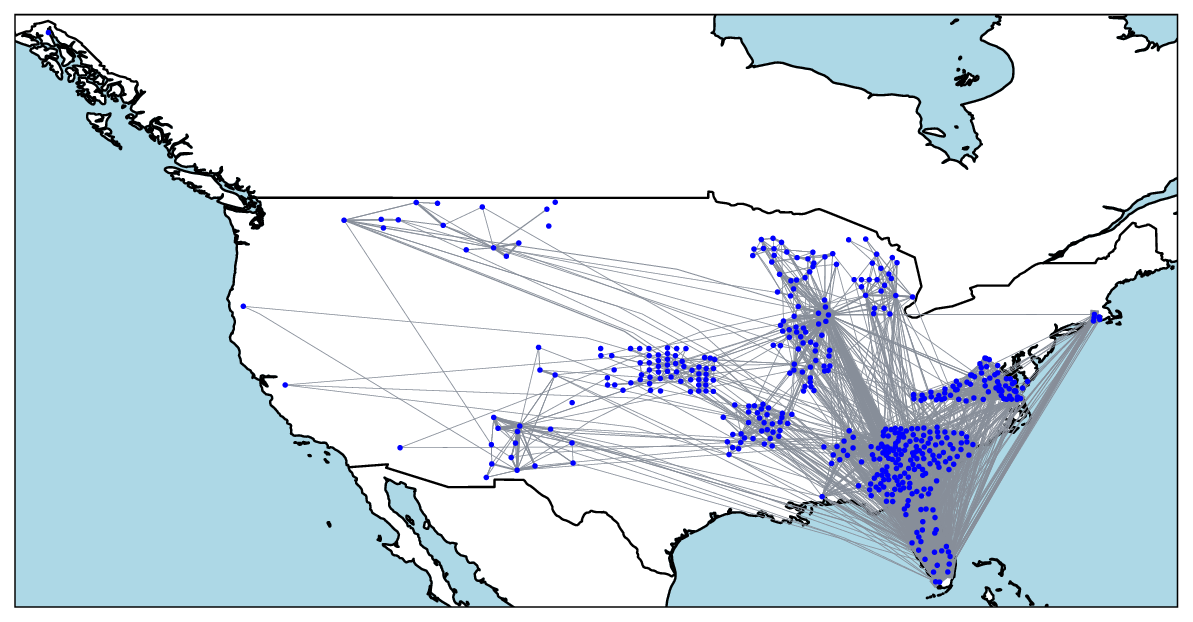}
    \caption{Visualization of the network of overdose journeys laid over a U.S. map. For visual clarity, the edge widths shown are proportional to the square root of the true edge weights. Self-loops and edge directionality are omitted.
    }
    \label{fig:network_visualization}
\end{figure}

Using EMS event records spanning January 2018 to March 2023, we construct a nationwide geospatial network from nonfatal opioid-involved overdose events. We aggregated the data on a semiannual basis, with each period representing six consecutive months and the first period of each year consisting of January through June. Each node of the network represents a county with a corresponding FIPS code. A directed edge ($u$, $v$) signifies an overdose journey from county $u$ to county $v$, and its edge weight represents the number of opioid-involved overdose events having occurred in county $v$, involving people who experienced an overdose residing in county $u$. Self-loops are edges where the source and destination counties are the same (i.e., $u=v$).  Edges that are not self-loops represent geographically discordant overdose events \citep{forati2023journey}. We show in Figure \ref{fig:network_visualization} the journey network after applying our exclusion criteria.

\subsection{Measures of networks}

\subsubsection{Degree}

The weighted in-degree of node $v$ is defined as the number of overdose events that originate from any other node and occur in node $v$. Similarly, the weighted out-degree of node $v$ is the number of overdose events originating from $v$ and occurring in a different node. In other words, the weighted in- and out-degree reflects the total number of overdose events imported from or exported to other counties. The unweighted in-degree of node $v$ is the number of other nodes for which an overdose event originates and occurs in node $v$, that is, the number of distinct counties of residence for people who experienced an overdose event occurring in $v$. The unweighted out-degree of node $v$ is the number of other nodes for which an overdose event originates in node $v$.

\subsubsection{Reciprocity of directed edges}

We measure the reciprocity of edges, denoted by $r$, which quantifies the tendency for nodes in a directed network to form mutual connections~\citep{newman2002email, garlaschelli2004patterns, newman2018networks}. It is defined by
\begin{equation}
r = \frac{{\sum_{i=1}^N \sum_{j=1}^N A_{ij} A_{ji}}}{{\sum_{i=1}^N \sum_{j=1}^N A_{ij}}},
\label{eq:def-r}
\end{equation}
where $N$ is the number of nodes, and $A=(A_{ij})$ is the unweighted adjacency matrix such that $A_{ij} = 1$ if there is an edge from the $i$th to the $j$th node, and $A_{ij} = 0$ otherwise. The numerator of Eq.~\eqref{eq:def-r} counts the number of reciprocal (i.e., bidirectional) edges, and the denominator counts the total number of directed edges in the network. The reciprocity ranges between $0$ and $1$.

\subsubsection{Hyperlink-induced topic search}

Hyperlink-Induced Topic Search (HITS), originally proposed in web analytics, ranks nodes in a directed network with respect to their roles as \textit{hubs} or \textit{authorities} \citep{kleinberg1999authoritative}. By definition, a hub is a node that tends to send outgoing edges to authority nodes, while an authority is a node that tends to receive incoming edges from hubs. A hub and authority are, at first glance, conceptually similar to nodes with a high out-degree and in-degree, respectively. However, a hub, for example, is different from a node with a high out-degree in that an $i$th node should send many outgoing edges to authority nodes, not ordinary nodes, for the $i$th node to become a strong hub.

The HITS algorithm assigns each $i$th node a hub score, denoted by $h(i)$, and an authority score, denoted by $a(i)$. These scores are initialized to 1 and follow an iterative update-normalize algorithm. A single updating round of the hub and authority score invokes the following formulae:
\begin{align}
h(i) =& \sum_{j \text{ such that } j \leftarrow i} a(j),
\label{eq:hub-updating}\\
a(i) =& \sum_{j \text{ such that } j \rightarrow i} h(j),
\label{eq:authority-updating}
\end{align}
after which we divide each $h(i)$ by $\sqrt{\sum_{j = 1}^N h(j)^2}$ and each $a(i)$ by
$\sqrt{\sum_{j = 1}^N a(j)^2}$ for normalization. 
We repeat these steps (i.e., 
Eqs.~\eqref{eq:hub-updating} and \eqref{eq:authority-updating}, followed by the normalization) until all $h(i)$ and $a(i)$ sufficiently converge. 

The hub score $h(i)$ quantifies the extent to which a node serves as a good ``recommender" of other nodes, aggregating the authority scores of the nodes to which the $i$th node points. Conversely, the authority score $a(i)$ aggregates the hub scores of all nodes pointing to $i$. By identifying hubs and authorities, i.e., nodes with high hub and authority scores, respectively, we aim to identify counties that are pivotal in the spatial distribution of opioid-involved overdoses, as was done in a previous study on overdose journeys \citep{forati2023journey}. Hub counties are interpreted to be focal exporters of opioid-involved overdose events, and authority counties are focal importers.

\subsubsection{Edge persistence}

In temporal networks represented by a time series of static networks in discrete time, edge persistence is the tendency for edges present at one time to be present in the subsequent time. We measure edge persistence by the temporal correlation coefficient~\citep{clausetpersistence, tang2010small, nicosia2013graph, Masuda2020book}. The usual formulation of the temporal correlation coefficient for the $i$th node, denoted by $\gamma_i$, is given by
\begin{equation}
\gamma_i = \frac{1}{T-1} \sum_{t=1}^{T-1} \frac{\sum_{j=1}^{N} A_{ij}(t) A_{ij}(t+1)}{\sqrt{\Big(\sum_{j=1}^{N} A_{ij}(t)\Big) \Big(\sum_{j=1}^{N} A_{ij}(t+1)\Big)]}},
\label{eq:gamma_i}
\end{equation}
where $N$ is the number of nodes in the temporal network, $T$ is the number of discrete time points, and $A(t)$ represents the $N \times N$ undirected adjacency matrix at time $t$, with $A_{ij}(t) \in \{ 0, 1 \}$. The network's overall temporal correlation coefficient is given by
\begin{equation}
\gamma = \frac{1}{N} \sum_{i=1}^{N} \gamma_i.
\end{equation}

As pointed out in \citep{buttner2016temporal}, Eq.~\eqref{eq:gamma_i} treats incoming edges and outgoing edges as interchangeable in the case of directed temporal networks. Therefore, we consider the in- and out- temporal correlation coefficients, denoted by $\gamma_i^{\text{in}}$ and $\gamma_i^{\text{out}}$, respectively, to focus on the persistence of incoming or outgoing edges. These temporal correlation coefficients for the $i$th node are defined by
\begin{equation}
\gamma_i^{\text{in}} = \frac{1}{T-1} \sum_{t=1}^{T-1} \frac{\sum_{j=1}^{N} A_{ji}(t) A_{ji}(t+1)}{\sqrt{\Big(\sum_{j=1}^{N} A_{ji}(t)\Big) \Big(\sum_{j=1}^{N} A_{ji}(t+1)\Big)}},
\end{equation}
\begin{equation}
\gamma_i^{\text{out}} = \frac{1}{T-1} \sum_{t=1}^{T-1} \frac{\sum_{j=1}^{N} A_{ij}(t) A_{ij}(t+1)}{\sqrt{\Big(\sum_{j=1}^{N} A_{ij}(t)\Big) \Big(\sum_{j=1}^{N} A_{ij}(t+1)\Big)}}.
\end{equation}

All temporal correlation coefficients described herein range between $0$ and $1$.

\subsection{Statistical methods}

\subsubsection{Estimating the distribution of overdose journey lengths}\label{sec:length_meth}

The accuracy of overdose journey distances is constrained by the censoring of granular location data (Section \ref{sec:ems}). To handle this constraint, we apply a method to reconstruct the distribution of transportation event distance \citep{mccabe2023nonparametric} to approximate the distributions of overdose journey lengths from FIPS codes. In particular, we carry out the following steps:
\begin{enumerate}
    \item \textbf{Boundary coordinate retrieval:} We obtain the latitude and longitude coordinates of points lying on the boundary of each FIPS code from the TIGER system, which provides shape files for counties.
    
    \item \textbf{Range estimation:} For each pair of FIPS codes, we calculate the shortest distance from any point on the boundary of the first to any point on the boundary of the second, according to ``as-the-crow-flies'' Haversine distance. The resulting range of distances provides a range within which a lower-bound distance between the overdose event and where the person who experienced an overdose lives.

    \item \textbf{CDF estimation:} We interpret the calculated distance ranges as interval-censored event records and estimate the corresponding survival function $S(d)$ using the Turnbull estimator  \citep{kaplan1958nonparametric, turnbull1976empirical}. The cumulative distribution function (CDF) is given by $F(d) = 1 - S(d)$.
\end{enumerate}

We apply two hypothesis-testing procedures to compare pairs of collections of censored overdose journey records (i.e., journeys of type A versus those of type B) in terms of their CDFs, denoted by $F_A(d)$ and $F_B(d)$, estimated in the previous step. We select these procedures because they mitigate the challenges associated with censored location data, and they do not rely on the non-guaranteed proportional hazards assumption common in hypothesis tests comparing survival functions. Of course, both procedures assume that the distributions of overdose journey lengths are well-estimated by steps 1-3 above.

In the first method, we employ the so-called Monte Carlo U-test for censored events, a stochastic dominance test for censored transportation event records based on an inverse transform sampling \citep{mccabe2023nonparametric}. In particular, we repeatedly draw samples from $F_A(d)$ and $F_B(d)$ and conduct Mann-Whitney U-tests for each sample \citep{mann1947test}. The p-values from the U-tests are then aggregated using Fisher's combined probability test \citep{fisherstatistical}. This approach inherits the assumptions of the U-test and Fisher's method: we assume independence of observations for the sampled journey lengths and independence among the U-tests.

In the second method, we apply a Z-test for the difference in the mean journey length. Specifically, we compute a Z statistic and its corresponding two-tailed $p$ value by $Z = (m_A - m_B)/\sqrt{(\sigma_A^2/n_A) + (\sigma_B^2/n_B)}$ and $p = 2 \Phi(-|Z|)$, where $m_A$, $\sigma_A$, and $n_A$ are the mean of $F_A(d)$, the standard deviation of $F_A(d)$, and the number of samples used for estimating $F_A(d)$, and similar for $m_B$, $\sigma_B$, and $n_B$. Function $\Phi$ denotes the CDF of the standard normal distribution. This approach is similar to the test of differences in so-called Restricted Mean Survival Time (RMST) found in clinical trials literature \citep{royston2011use, royston2013restricted}. Although the true mean and standard deviation of journey lengths can be theoretically obtainable via analytical integration the true CDFs, they are not known. This test relies on the assumption that $m_A$, $\sigma_A$, $m_B$, and $\sigma_B$ are well-estimated via numerical integration of $F_A(d)$ and $F_B(d)$.

\subsubsection{Other statistical analyses}

We employ several hypothesis tests other than the U- and Z-tests described in Section \ref{sec:length_meth}. For group comparisons based on contingency tables, we apply the G-test, which is a log-likelihood ratio-based goodness-of-fit test \citep{mcdonald2014g}. It is an alternative to the more common chi-squared test, but is often recommended instead on theoretical and practical grounds \citep{sokal1987biostatistics}. When assessing differences in means, we invoke Tukey's honestly significant difference (HSD) test. We choose the HSD test instead of alternatives involving analysis of variance with subsequent pairwise comparisons, because HSD automatically corrects for multiple comparisons \citep{tukey1949comparing, kramer1956extension, abdi2010tukey}. When testing the significance of a trend of a time series, we conduct the Hamed and Rao trend test \citep{hamed1998modified} over the original Mann-Kendall method \citep{kendallrank} because the former test accounts for the effects of autocorrelation. 
When performing multiple comparisons with tests that do not automatically account for them, we apply the Holm-Bonferroni correction to control family-wise error rate \citep{holm1979simple}. This procedure ensures that the false positive rate does not exceed the significance level, while maintaining a lower false negative rate than the Bonferroni adjustment \citep{abdi2010holm}.

\hfill \break

The activities described in this paper were reviewed by CDC and conducted consistently with applicable federal law and CDC policy.

\section{Results\label{sec:nationwide}}

\subsection{Basic properties}\label{sec:basic}

After applying our exclusion criteria (Section \ref{sec:ems}), we are left with $N=$ 481 counties from $19$ states. We show basic structural properties of the overdose journey network in Table \ref{tab:basicstats}. The overwhelming majority of events (93.7\%, 467096 events) occur within the FIPS code of residence of the person who experienced an overdose, represented by self-loops. However, approximately 6.3\% (31385 events) occur outside the FIPS code of residence of the person who experienced an overdose. Such events are geographically discordant and are the focus of this work.  Of these geographically discordant overdose events, 70\% occurred between geographically adjacent counties, and 30\% between non-adjacent counties.

\begin{table}[htbp]
    \centering
        \caption{Basic network properties of the overdose journey network. ``Included" and ``Excluded" refer to the networks including and excluding the self-loops, respectively.}
    \begin{tabular}{|l|c|c|}
        \hline
        Self-loops & Included & Excluded \\
        \hline
        Number of nodes                 & 482     & 481     \\
        Number of edges                 & 4845    & 4363    \\
        Maximum weighted in-degree      & 74612   & 1667    \\
        Maximum weighted out-degree     & 73533   & 1000     \\
        Mean weighted in/out-degree     & 1034.19 & 65.25   \\
        Maximum unweighted in-degree    & 106   & 105    \\
        Maximum unweighted out-degree   & 61   & 60     \\
        Mean unweighted in/out-degree   & 10.05 & 9.07   \\
        Reciprocity                     & 0.436   & 0.484 \\
        \hline
    \end{tabular}
    \label{tab:basicstats}
\end{table}

\subsection{Degree analysis}\label{sec:degree_analysis}

\begin{figure}[t!]
    \centering
    \includegraphics[width=1.0\textwidth]{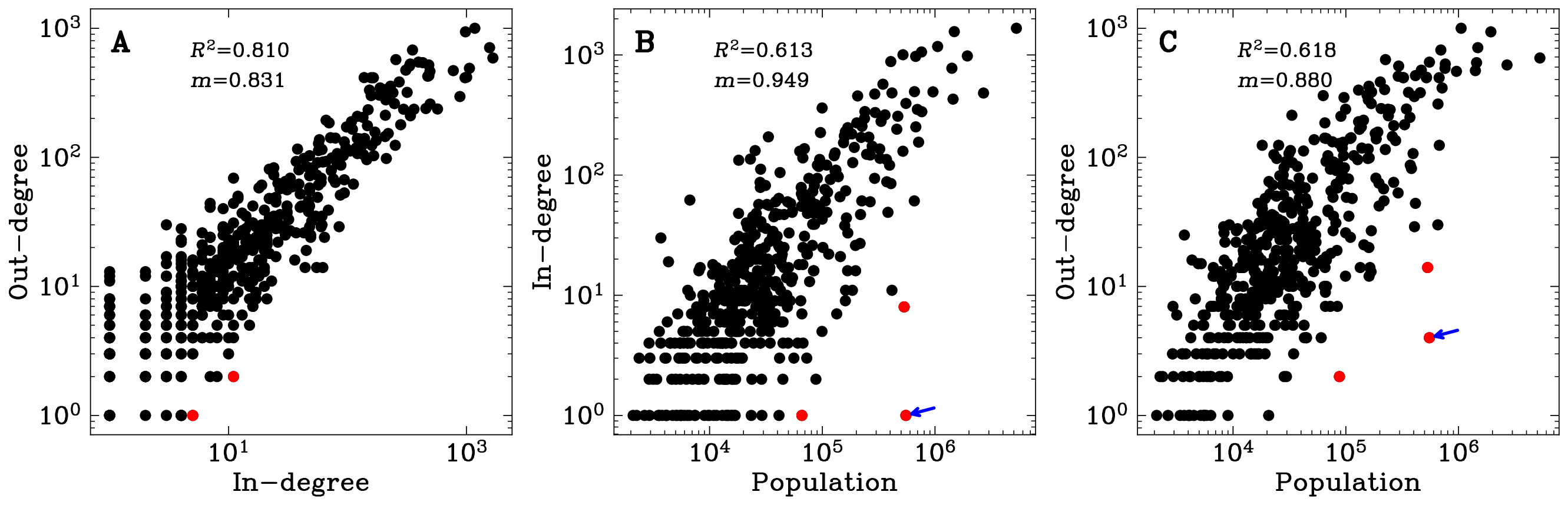}
    \caption{Relationships between the weighted in-degree, weighted out-degree, and population of the node. \textbf{A)} Relationships between the weighted in-degree and the weighted out-degree in the overdose journey network excluding self-loops. \textbf{B)} Relationship between the weighted out-degree and the population. \textbf{C)} Relationships between the weighted in-degree and the population. In-degree and out-degree values have been incremented by one to ensure visibility in log-transformed space. 
    In all three cases, the weighted degrees exhibit slightly sub-linear relationships as functions of population size.
    The $R^2$ values represent correlations between log-transformed values. The $m$ values represent the slopes of the linear regression in the log-transformed space. Outlier points, i.e., counties for which the magnitude of the residuals between observed values and those predicted by the linear regression exceed three standard deviations, are highlighted in red. The blue arrows indicate the county that is an outlier in both panels B and C.}
    \label{fig:degree_corr_weighted}
\end{figure}

We investigate relationships between the weighted in-degree, weighted out-degree, and the population. Because the network degrees and city populations often obey heavy-tailed distributions \citep{clauset2009power}, we take the logarithm of these three quantities and show the scattergram and $R^2$ between each pair of the quantities in Figure \ref{fig:degree_corr_weighted}. The figure indicates that these three quantities are strongly positively correlated with each other. We also observe slightly sub-linear relationships for the weighted degrees as a function of the population, which suggests a saturation effect of high-population counties as destinations and sources of overdose behavior (Figure \ref{fig:degree_corr_weighted}B and C). These results are qualitatively the same for unweighted in- and out-degrees, although the correlation is weaker (see Figure \ref{fig:degree_corr_unweighted}). 

We define outlier points as those whose absolute value of the residual obtained from the linear regression in the log-transformed space exceeds three standard deviations. We show the outlier points in red in Figure \ref{fig:degree_corr_weighted}. Of particular interest is the one county that is an outlier in both Figure \ref{fig:degree_corr_weighted}B and C (indicated with a blue arrow), representing disproportionately low geographically discordant overdose activity. This county is a particularly Hispanic and Latino medium metropolitan area that serves as a commuter county for three nearby large metropolitan areas.

\begin{table}[b!]
    \centering
        \caption{Urbanization, demographic, and socioeconomic profiles of the top ten counties by in-degree per capita (top IPC) and out-degree per capita (top OPC), compared to the remaining counties. AI/AN stands for American Indian and Alaska Native. NHPI stands for Native Hawaiian and other Pacific Islander. 
        The top IPC and OPC counties are exclusively urban.
        Poverty refers to the fraction of the population living below poverty level. The urbanicity values are not population-weighted. Urbanization categories are collapsed into urban and rural per Section \ref{other_data}. The corresponding results for the uncollapsed urbanization categories are shown in Table \ref{tab:degree_demographics_uncollapsed}.
        }
    \begin{tabular}{|l|l|>{\centering\arraybackslash}p{0.15\linewidth}|>{\centering\arraybackslash}p{0.15\linewidth}|>{\centering\arraybackslash}p{0.15\linewidth}|}
    \hline 
         & Category & \textbf{Top IPC} & \textbf{Top OPC} & \textbf{Other} \\
    \hline
        \textbf{Population} & Mean Population & 1384559 & 1370947 & 70262 \\
        \hline
        \textbf{Urbanicity} & Urban & 100.0 \% & 100.0 \% & 30.57 \% \\
        & Rural & 0.0 \% & 0.0 \% & 60.43 \% \\
    \hline
        \textbf{Race and} & White Alone & 53.88 \% & 55.66 \% & 71.27 \% \\
        \textbf{Ethnicity} & Black Alone & 26.83 \% & 24.21 \% & 15.61 \% \\
        & AI/AN Alone & 0.352 \% & 0.331\% & 0.900 \% \\
        & Asian Alone & 5.365 \% & 5.579 \% & 2.208 \% \\
        & NHPI Alone & 0.056 \% & 0.048 \% & 0.075 \% \\
        & Hispanic or Latino & 21.72 \% & 23.09 \% & 16.54 \% \\
    \hline
        \textbf{Economic} & Employed & 52.84 \% & 52.68 \% & 48.25 \% \\
        & Poverty & 12.90 \% & 12.30 \% & 13.10 \% \\
    \hline
    \end{tabular}
    \label{tab:degree_demographics}
\end{table}

To better understand focal points of geographically discordant overdose activity, we consider the top ten counties by the weighted in-degree per capita (top IPC counties) and top ten counties by the weighted out-degree per capita (top OPC counties), representing disproportionately high imports and exports of opioid overdoses when normalized by county population. Even after adjusting for population size, these top counties are exclusively urban counties (Table \ref{tab:degree_demographics}), but they do not simply reflect the largest cities. The top IPC counties are roughly equally represented across large central metro, large fringe metro, and medium metro areas, whereas the top OPC counties are concentrated in large central metro and large fringe metro areas (Table \ref{tab:degree_demographics_uncollapsed}). Notably, the top IPC and top OPC counties have populations an order of magnitude larger and significantly higher employment rates than counties that are neither. This may be explained by differences in urbanicity alone: cities are larger and have more jobs than non-cities. The group difference between the top IPC counties and the others in terms of poverty rate is not statistically significant ($p =$ 0.139), nor is the difference between the top IPC and top OPC counties in terms of population size ($p=0.995$). All other group differences are significant ($p < 0.05$). The test results are provided in Table \ref{tab:sig_ipcsopcs_hits}.

\subsection{HITS analysis}\label{sec:hits}

We similarly characterize in Table \ref{tab:hits_demographics} the counties with the highest authority and hub scores in comparison to all other counties. Hubs and authorities are over-represented in large fringe metro areas (Table \ref{tab:hits_demographics_uncollapsed}). However, despite its name, a large fringe metro county in a large MSA may have a relatively small population because urban-rural classifications depend on MSA populations. In fact, on average, while the top authorities have approximately 12\% larger populations than the top IPC and OPC counties, the top hubs have roughly 61\% the populations of the top
authorities and 69\% that of the top IPC and OPC counties (see Table \ref{tab:hits_demographics}). This result is in stark contrast with the observation that the mean population is only approximately 1\% different between the top IPC and the top OPC counties (Table \ref{tab:degree_demographics}). 
Additionally, hub counties have a lower poverty rate than the top authorities or the other counties. All reported group differences are significant with $p <$ 0.01. The test results are provided in Table \ref{tab:sig_ipcsopcs_hits}.

\begin{table}[htbp]
    \centering
    \caption{Urbanization, demographic, and socioeconomic profiles of the top ten and authorities and hubs, compared to the remaining counties. AI/AN stands for American Indian and Alaska Native. NHPI stands for Native Hawaiian and other Pacific Islander. 
    Like the top IPC and OPC counties, the top hub and authority counties are exclusively urban. Hubs have significantly smaller average populations than authority, top IPC, or top OPC counties.
    Poverty refers to the fraction of the population living below poverty level. The urbanicity values are not population-weighted. Urbanization categories are collapsed into urban and rural per Section \ref{other_data}. The corresponding results for the uncollapsed urbanization categories are shown in Table \ref{tab:hits_demographics_uncollapsed}.
    }
    \begin{tabular}{|l|l|>{\centering\arraybackslash}p{0.15\linewidth}|>{\centering\arraybackslash}p{0.15\linewidth}|>{\centering\arraybackslash}p{0.15\linewidth}|}
    \hline
         & Category & 
         \textbf{Authority} & \textbf{Hub} & \textbf{Other} \\
    \hline
    \textbf{Population} & Mean Population & 1552178 & 952128 & 71659 \\
    \hline
    \textbf{Urbanicity} & Urban & 100.0 \% & 100.0 \% & 39.96 \% \\
    & Rural & 0.0 \% & 0.0 \% & 60.04 \% \\
    \hline
        \textbf{Race and} & White Alone & 55.83 \% & 59.15 \% & 70.98 \% \\
        \textbf{Ethnicity} & Black Alone & 21.43 \% & 20.35 \% & 17.28 \% \\
        & AI/AN Alone & 0.367 \% & 0.290 \% & 0.878 \% \\
        & Asian Alone & 5.086 \% & 3.821 \% & 2.270 \% \\
        & NHPI Alone & 0.046 \% & 0.036 \% & 0.0816 \% \\
        & Hispanic or Latino & 32.94 \% & 35.00 \% & 11.96 \% \\
    \hline
        \textbf{Economic} & Employed & 52.96 \% & 52.60 \% & 48.13 \% \\
        & Poverty & 12.62 \% & 12.13 \% & 13.03 \% \\
    \hline
    \end{tabular}
    \label{tab:hits_demographics}
\end{table}

\subsection{Lengths of geographically discordant overdose journeys}\label{sec:journeylength}

\begin{figure}[h]
    \centering
    \includegraphics[width=1.0\textwidth]{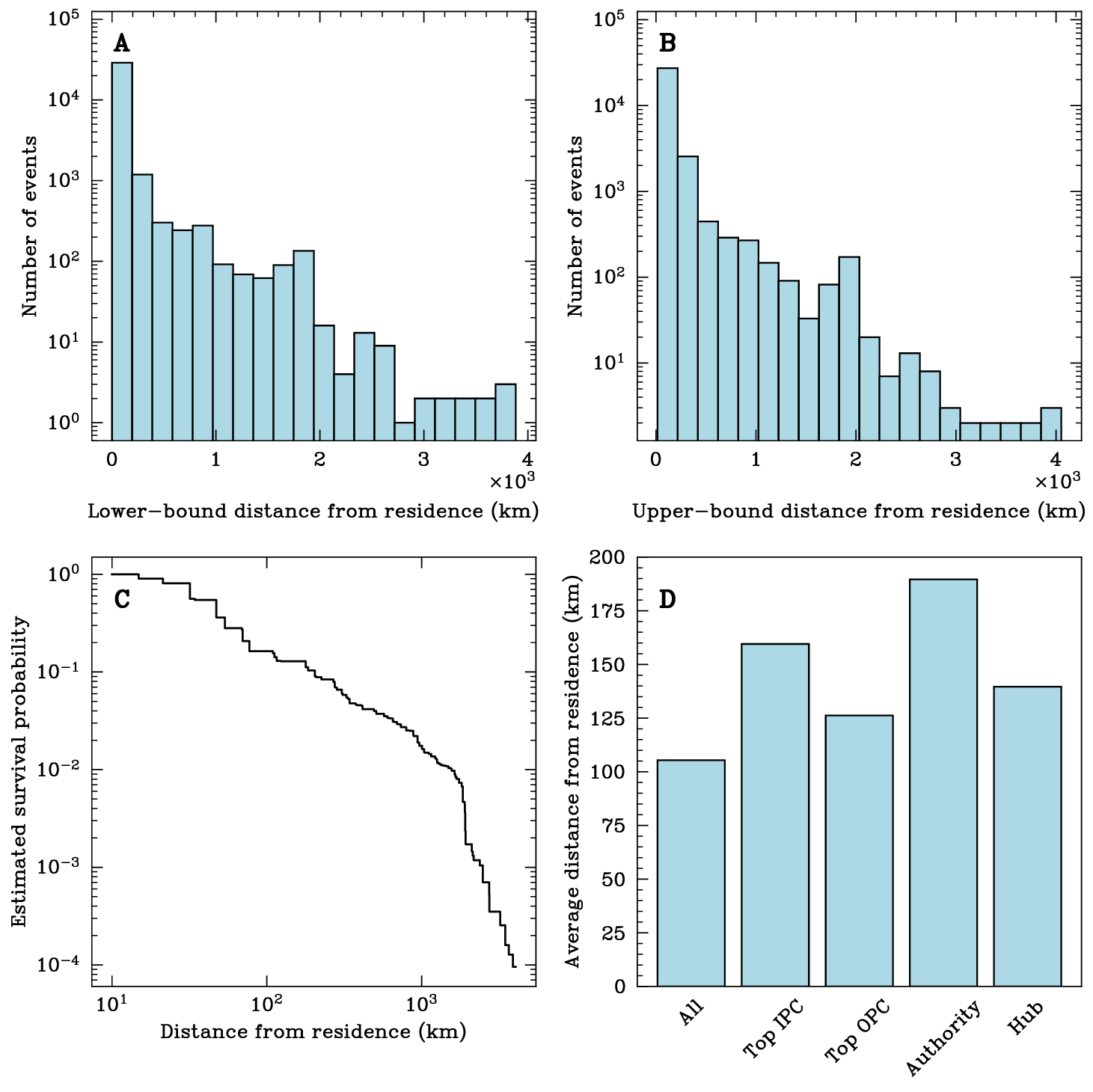}
    \caption{Distribution of distances of geographically discordant overdose journeys. \textbf{A)} Empirical distribution of lower-bound distances among the journeys.  \textbf{B)} Empirical distribution of upper-bound distances. \textbf{C)} Estimated survival function of the journey distance. \textbf{D)} Average distance of geographically discordant overdose journeys. ``All'' refers to all journeys. ``Top IPC'' and ``Authority'' refer to journeys to the top IPC and authority counties, respectively. ``Top OPC'' and ``Hub'' refer to journeys from the top OPC and hub counties, respectively.
    }
    \label{fig:distances_static}
\end{figure}

Figure \ref{fig:distances_static} shows the distribution of journey-to-overdose lengths, denoted by $d$, among the geographically discordant overdose events. The histogram mass to the right of a particular lower-bound distance $d$ represents the number of events recorded at least $d$ (Figure \ref{fig:distances_static}A). Similarly, the histogram mass to the left of a particular upper-bound distance $d$ represents the number of events recorded at most $d$ (Figure \ref{fig:distances_static}B). From these lower-bound and upper-bound distances, we estimate the distribution of journey distances as described in Section \ref{sec:length_meth}. Figure \ref{fig:distances_static}C displays the resultant estimated complementary CDF (i.e., survival function) of $d$.

From our estimation, the median and mean distance of geographically discordant overdose journeys is 47.54 km and 105.4 km, respectively (Figure \ref{fig:distances_static}D). Unsurprisingly, events are more common close to the home of a person who experienced an overdose. However, approximately 10\% of the journeys exceed 204.38 km, and 5\% exceed 342.56 km. Figure \ref{fig:distances_static}C indicates a heavy-tailed nature of the distribution. Indeed, the coefficient of variation (CV) of $d$, i.e., the standard deviation of $d$ divided by the mean of $d$, is substantially larger than $1$ ($\text{CV}=$ 2.330), indicating that the distribution of $d$ has a heavier tail than the exponential distribution.
 
In Figure \ref{fig:distances_static}D, we interpret the journey-to-overdose distance as an import or export radius. In particular, we investigate the average journey-to-overdose distance \textit{to} importers for the top IPC and authority counties and \textit{from} exporters for the top OPC and hub counties. As reference, we also show the journey distance averaged over all the journeys (see the ``All'' bar in Figure \ref{fig:distances_static}D). The geographically discordant overdose journey from the residence to the top IPC counties and from the top OPC counties to their destinations are both longer, on average, than the mean journey. Furthermore, the journey distance for the top IPC counties is larger than that for the top OPC counties on average. As expected, the average journey distance to the top authority counties and  from the top hub counties show qualitatively the same behavior as that to the top IPC counties and that from the top OPC counties, respectively. In other words, the average distance to the top authority counties is larger than that from the top hub counties, and the latter is larger than the average over all geographically discordant journeys. However, the effects of authorities and hubs are not a straightforward consequence of the general fact that authorities and hubs tend to be counties with large in-degrees and large out-degrees, respectively. In fact, Figure~\ref{fig:distances_static}D suggests that the average journey distance is larger for the top authority counties than the top IPC counties and larger for the top hub counties than the top OPC counties. See the ``Monte Carlo U-Test" and ``Z-Test" columns of Table \ref{tab:sig_dist_corr} for statistical results.

\subsection{Temporal aspects}\label{sec:temporal}

\begin{figure}[h!]
    \centering
    \includegraphics[width=1.0\textwidth]{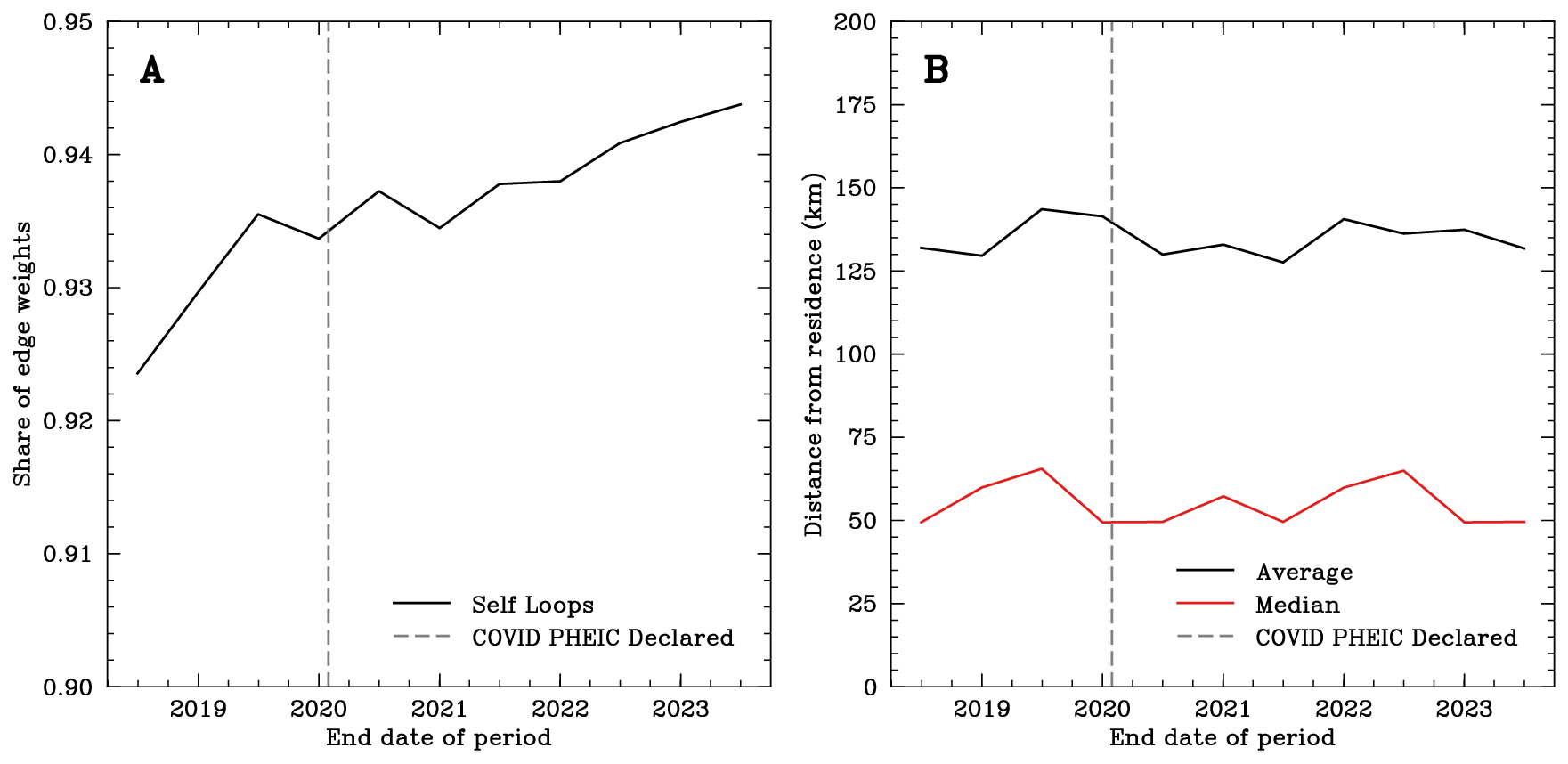}
    \caption{Time courses of the fraction of self-loops and the overdose journey distance. \textbf{A)} Fraction of edge weights owing to self-loops, indicating the portion of overdose events occurring in the home county of the person who experienced an overdose.
    \textbf{B)} Average and median journey distance for geographically discordant events.
    The dashed lines indicate the date at which the World Health Organization declared the beginning of the COVID-19 Public Health Emergency of International Concern (PHEIC) \citep{npr_covid_2023}.
    }
    \label{fig:dists_ts}
\end{figure}

We decompose the overdose journey network into a time series of overdose journey networks by constructing the network for each six-month time window. 
We show the time series of the share of the self-loops and the average journey distance in Figure \ref{fig:dists_ts}A and B, respectively, in which each data point represents a six-month time window.
The portion of the overdose events occurring in the person's home county increased throughout 2018 to 2023 (Figure \ref{fig:dists_ts}A). 
The average and median journey distance for geographically discordant overdose events did not systematically increase or decrease over the same five years, including the wake of the COVID-19 pandemic, which is shown by the dashed line (Figure \ref{fig:dists_ts}B).

\begin{figure}[h!]
    \centering
    \includegraphics[width=1.0\textwidth]{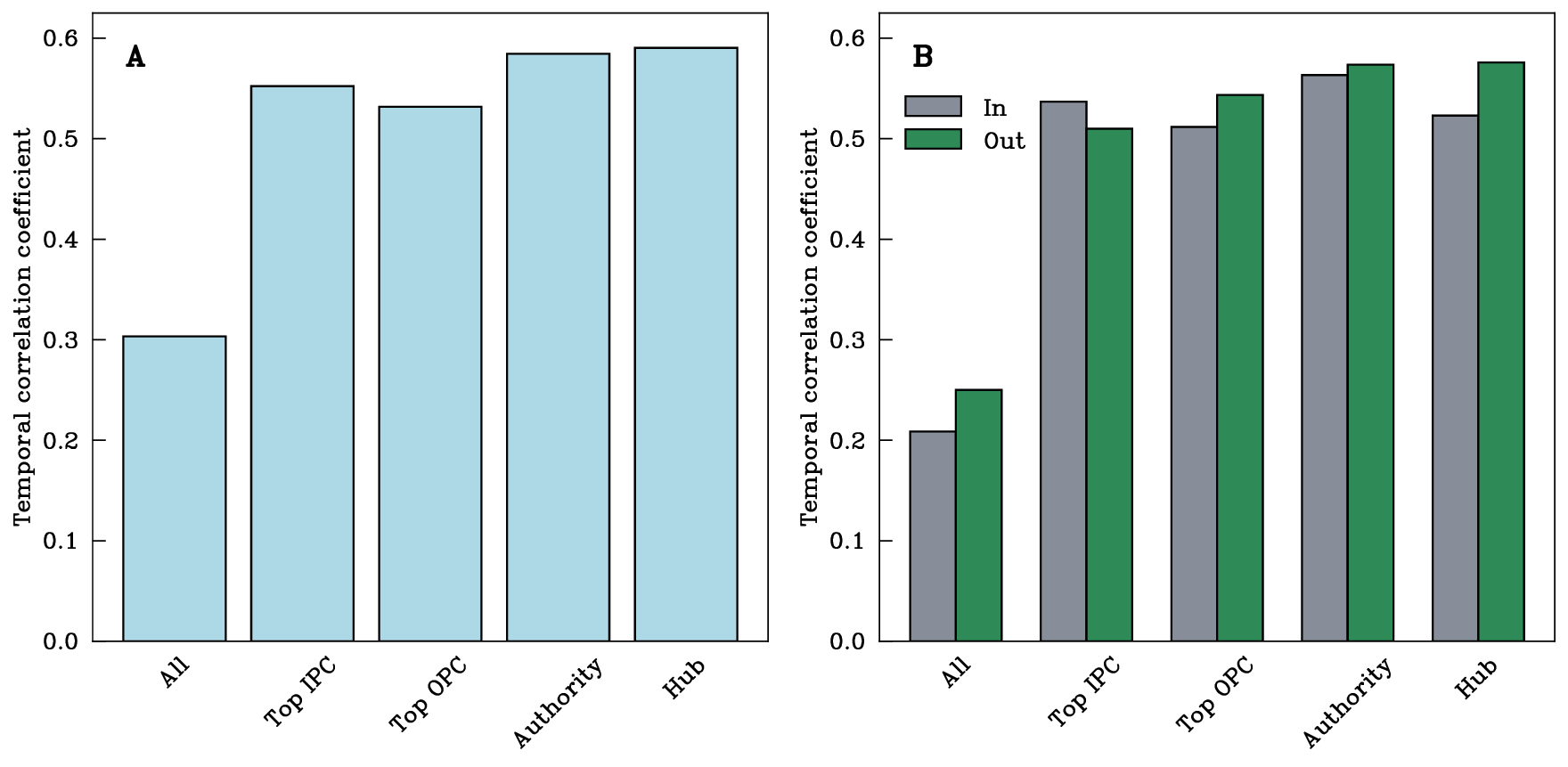}
    \caption{Persistence of edges in the temporal overdose journey network. 
    \textbf{A)} Average undirected temporal correlation coefficients for all counties, and the top IPC, OPC, authority, and hub counties.
    \textbf{B)} Average directed (i.e., in- and out-) temporal correlation coefficients for all counties, and the top IPC, OPC, authority, and hub counties. 
    In both the undirected and directed settings, focal importers and exporters have greater-than-average temporal correlation coefficients.
    }
    \label{fig:t_corr}
\end{figure}

Whether edges of the journey network are consistently present from one time period to another is relevant for public health applications. Such persistence reflects enduring overdose-related transit patterns that may warrant policy intervention. To quantify the persistence of edges, we obtain the undirected temporal correlation coefficient $\gamma =$ 0.303. Further, focal importers and exporters (i.e., the top IPC, OPC, authority, and hub counties) exhibit a significantly greater probability of (undirected) incident edge persistence than the average over all counties ($p < 0.05$), with $\gamma$ nearly double the network's average (Figure \ref{fig:t_corr}A). Approximately the same results hold true in the directed setting, as well: focal importers and exporters also have significantly greater in- and out- temporal correlation coefficients than the network's average ($p < 0.05$, Figure \ref{fig:t_corr}B).

\section{Discussion}\label{sec:conclusions}

We analyzed a spatial network of opioid-involved overdose journeys from $19$ states across the U.S. constructed from EMS records spanning 2018--2023. Our work underscores that the opioid overdose epidemic does not only exist where overdoses occur. Patterns of import and export of overdose behavior across county borders emerge, as more than 6\% of opioid-involved overdose events occurred outside the residential FIPS code of the person who experienced an overdose. We focused on these geographically discordant events, of which roughly $10\%$ occurred over 204 km from the home of the person who experienced an overdose.
It is itself noteworthy, however, that nearly 94\% of opioid-involved overdoses occurred within the county of residence of the person who experienced the overdose, suggesting that the majority of persons overdosed in their home communities. Thus, assessment of either the location of the overdose event or the documented patient residence can be used to understand focal locations for overdoses, such that harm reduction services (e.g., naloxone and fentanyl test strip distribution) and services for linking to care and treatment can be deployed more effectively.

Public health authorities may wish to identify and intervene at focal points of import/export behavior, but it is not necessarily clear how to define them. In this work, we considered two such approaches: measures calculable from normalized travel counts involving the county alone (i.e., IPC and OPC) and network-based measures (i.e., authorities and hubs). Focal counties according to these measures tended to be urban, were higher in Black, Asian, and Hispanic or Latino populations, were lower in white, American Indian and Alaska Native, and Native Hawaiian and other Pacific Islander populations, and had a larger employment rate, population, journey distance, and edge persistence than the average over all the edges in the network.

We found that some aspects of overdose journeys differed between the types of focal points. For instance, the top IPC, OPC, and authority counties have population sizes of the same order of magnitude on average, but the top hubs are much smaller in population, although they are much larger than the average over all counties. Additionally, the overdose journeys to the top IPC and authority counties were longer than those from the top OPC and hub counties on average.
This result may be partly explained by size and resource advantage: the average population of authorities is more than twice that of hubs, and larger counties may provide more resources for opioid users \citep{chatterjee2022broadening}.

We contend that indirect paths in the overdose journey network do not have functional meanings for the following reason: If there is an indirect path from the $i$th to the $\ell$th nodes, then $A_{ij} = 1$ and $A_{j\ell}=1$, which implies that there is an overdose journey from the $i$th to the $j$th nodes, made by individual $u_1$, and another journey from the $j$th to the $\ell$th nodes, made by individual $u_2$. In this situation, it is unlikely that $u_1$ influences $u_2$ to trigger $u_2$'s overdose behavior outside $u_2$'s residence. The same reasoning holds true even if edges are weighted or we consider the effects of mass media or social media. The difficulty of interpretation of indirect paths may limit the utility of network analysis as an approach, rendering only the direct path effects such as the in- and out-degrees relevant to overdose behavior. However, we found that the HITS algorithm, which exploits effects of the indirect paths (e.g., hub scores of nodes are large when they send edges to authorities, which are nodes that tend to receive edges from hubs), reveals features of overdose journeys that were beyond the prediction by the top IPC or OPC counties. Specifically, the top authority and hub counties had larger journey distances than the top IPC and OPC counties, respectively. In addition, the top authorities, top hubs, and top OPC counties tend to be concentrated in large fringe metropolitan areas, whereas top IPC counties are more dispersed across large central metropolitan, large fringe metropolitan, and medium metropolitan areas.
These results suggest that the authorities and hubs, which tend to be interconnected with each other, may form a scaffold of the overdose journey network and have particular properties compared to top IPC and OPC counties. Further investigating this issue in combination with other data sources, demographic information, and geographic information, with potentially improved coverage and accuracy of the EMS records, warrants future work.

We showed in Section~\ref{sec:temporal} that the share of geographically discordant events decreased between 2018 and 2023. This result is consistent with recent evidence from Rhode Island suggesting that drug overdose deaths occurring in residences of the people who experienced overdoses increased disproportionately from $2019$ to $2020$ \citep{macmadu2021comparison}.
Among geographically discordant events, however, we did not observe a decrease in overdose journey distance after the beginning of the COVID-19 pandemic. 
Future work examining the individual-level profiles of this type of overdose journey would enhance this line of inquiry.

Our estimate of the fraction of geographically discordant overdose events is much lower than the 25\% reported in a recent study \citep{forati2023journey}. There are at least two potential reasons causing this difference. First, their work examined overdose deaths, whereas we focused on nonfatal incidents.
Second, we use differently sized geographic units: counties in the present study versus census tracts in their study. Therefore, our criteria for geographical discordance is technically stricter than theirs, excluding many events that could be considered discordant by the definition used in their work.
In addition to these differences, our larger scale of analysis enabled us to identify inter-county overdose journeys and analyze the distribution of overdose journey distances. Another advancement of the present study relative to theirs is that we compared between authority counties and top IPC counties and between hub counties and top OPC counties. With the latter analysis, we provided evidence supporting that the HITS algorithm provides unique information about overdose journeys that simpler pairwise analysis does not.

Our work has at least four potential limitations. 
First, we are limited by data availability because we have few or no records for many states in the U.S. This may introduce bias in our results and limit generalizability, since the availability of EMS data may be influenced by latent confounders. That is, our end-to-end selection process may not have resulted in a uniformly random subset of all nonfatal opioid-involved overdose events. Furthermore, U.S. demographics have significant spatial variability, and many aspects of the opioid epidemic have racial covariates \citep{santoro2018racial, gondre2023opioid}, which probably contribute to heterogeneity in overdose journey statistics across states.

Second, related to the first limitation, we restricted our analysis to the counties with at least 75\% UEC for the entire five-year period. Doing so is standard for the present data set \citep{casillas2022patient} but excludes many counties (Section \ref{sec:ems}). Future data collection efforts may mitigate this limitation, as the number of reporting counties increases. 
Third, we focused on non-fatal opioid-involved overdose events, which may mean our characterizations of journey lengths is not fully representative of high-mortality counties. 
Fourth, our EMS event records are geographically censored at the county-equivalent level. Therefore, we make no claims about spatial dynamics for within-county overdose journeys. Additionally, we rely on survival analysis-based procedures for estimating the distributions of overdose journey lengths. Since these approaches are limited by the quality of survival function estimation, future work may aim to narrow these data gaps to draw more granular insights regarding the national character of overdose journeys. 
Future work can also focus on different types of drugs, such as stimulants, to compare the spatial dynamics of different non-fatal overdoses across the country.
In the U.S., the overdose death rate involving psychostimulants with abuse potential (e.g., methamphetamine) has historically been higher in rural counties than urban ones \citep{spencer2022urban}. We are not aware of any work examining network patterns of stimulant-involving overdose journeys.
Further, more comprehensive analysis of socioeconomic variables may provide a richer characterization of overdose journeys. Of particular interest are proxies of healthcare access, such as primary care physicians per unit population \citep{CHRR2024} or response to related questions in the CDC's Behavioral Risk Factor Surveillance System
\citep{silva2014behavioral, buchmueller2020aca}.

We highlight the value of leveraging large-scale EMS data for understanding the spatial dynamics of the opioid crisis. 
While spatially aggregated EMS records allow researchers to conduct analyses while preserving patient privacy, there are important limitations arising from data availability and completeness. These may be partially ameliorated as coverage and reporting rates steadily rise, but this may take a long time. Future work should consider if integration of external and partially redundant data sources can support analyses such as the present study with lower minimum UEC levels.
A second potential avenue for future research is to leverage agent-based simulation, which has already been used to study spatio-temporal dynamics of opioid use disorder \citep{grefenstette2013fred}, to better understand the efficacy of targeting authorities and hubs for public health interventions. 
Of particular interest is the relationship between the number and strategic placement of harm-reduction and emergency room facilities and overdose journey length. Ambulance journey distance has been shown to be correlated with patient mortality \citep{nicholl2007relationship}, so a function of overdose journey length may be a reasonable proxy for the likelihood of overdose survival in such simulations.
Third, further analysis contrasting self-loops and discordant overdose journeys may be fruitful. Such an analysis can involve a review of the type of locations where these events occur (e.g., private residences, public areas such as streets, schools, etc.) or patient demographic characteristics, which may provide more insight into the differences among those who overdosed in their home communities and those who overdosed elsewhere. Then, more tailored public health interventions may be deployed.
A fourth future research thrust is to better understand the especially long overdose journeys highlighted in this study. In Section \ref{sec:journeylength}, we estimate that 5\% of geographically discordant incidents occurred over 426 km from the home of the person who experienced an overdose. We emphasize, however, that people who experienced overdoses did not necessarily travel such distances on the days of their incidents. For instance, not all individuals (e.g., those experiencing homelessness) functionally reside in their official residence of record for the entire year. Additionally, EMS place-of-residence information is sometimes collected from historical patient records, which may be out-of-date or inaccurate due to clerical errors. Further research is required to better understand the etiology of these particularly long overdose journeys.

\section*{Declarations}
\subsection*{Availability of data and materials}

Data were provided by biospatial, Inc. through an existing data use agreement with the Centers for Disease Control and Prevention, National Center for Injury Prevention and Control. Due to the existing clauses of the data use agreement, we do not make the data available for public release. The list of hub and authority counties, however, is accessible at reasonable request of the authors.

\subsection*{Acknowledgements}

We thank Freelancer International Pty Limited for programmatic support and Josh Walters for assistance involving the biospatial platform.

\subsection*{Authors' contributions}

LHM, NM, ND, and RL, conceptualized the study. LHM, NM, and SC provided the methods.  LHM implemented the methods, analyzed the data, and generated the visualizations. LHM and NM mainly wrote the manuscript, with contributions from SC, ND, and RL. All authors discussed the methods and results and reviewed the manuscript.

\subsection*{Funding}

This material is based upon work supported by the Centers for Disease Control and Prevention under Contract No. NOIS2-096. SC, AA, and RL are employed by the Centers for Disease Control and Prevention, National Center for Injury Prevention and Control, which also funded this work.

\subsection*{Competing interests}

The authors declare that they have no competing interests.

\begin{appendices}

\clearpage
\section{}\label{sec:supplementary}

\renewcommand{\thefigure}{\Alph{section}\arabic{figure}}
\renewcommand{\thetable}{\Alph{section}\arabic{table}}
\setcounter{figure}{0}
\setcounter{table}{0}

\subsection{Acronyms used in this work\label{sec:acronyms}}

\begin{table}[h]
\centering
\begin{tabular}{@{}lp{0.25\linewidth}p{0.55\linewidth}@{}}
\toprule
\textbf{Acronym} & \textbf{Full Form} & \textbf{Description} \\ \midrule
ACS              & American Community Survey & Annual demographics survey of approximately 3.5 million U.S. addresses managed by the U.S. Census, providing details on education, employment, housing, and more. \\
CSTE             & Council of State and Territorial Epidemiologists & U.S. non-profit public health organization. \\
EMS              & Emergency Medical Services & A healthcare system offering first-respondent care to medical emergencies. \\
FIPS             & Federal Information Processing Standards & Numerical codes assigned by the National Institute of Standards and Technology (NIST) to uniquely represent U.S. states and counties (including county equivalents). \\
NCHS             & National Center for Health Statistics & Agency of the U.S. Centers for Disease Control and Prevention (CDC) providing public health statistical information. \\
UEC              & Underlying Event Coverage & Proprietary metric by biospatial estimating the portion of total EMS events in a given area captured in the biospatial platform for a given time period. \\
URCS             & Urban–Rural Classification Scheme for Counties & Classification system provided by the U.S. National Center for Health Statistics (NCHS) assigning each U.S. county to one of six urban categories. \\
TIGER             & Topologically Integrated Geographic Encoding and Referencing & System maintained by the U.S. Census providing digital cartographic information of U.S. geographic areas. \\
\bottomrule
\end{tabular}
\caption{A selection of acronyms used in this article.}
\label{tab:acronyms}
\end{table}

\subsection{Distribution of UEC}

\begin{figure}[H]
    \centering
    \includegraphics[width=0.99\textwidth]{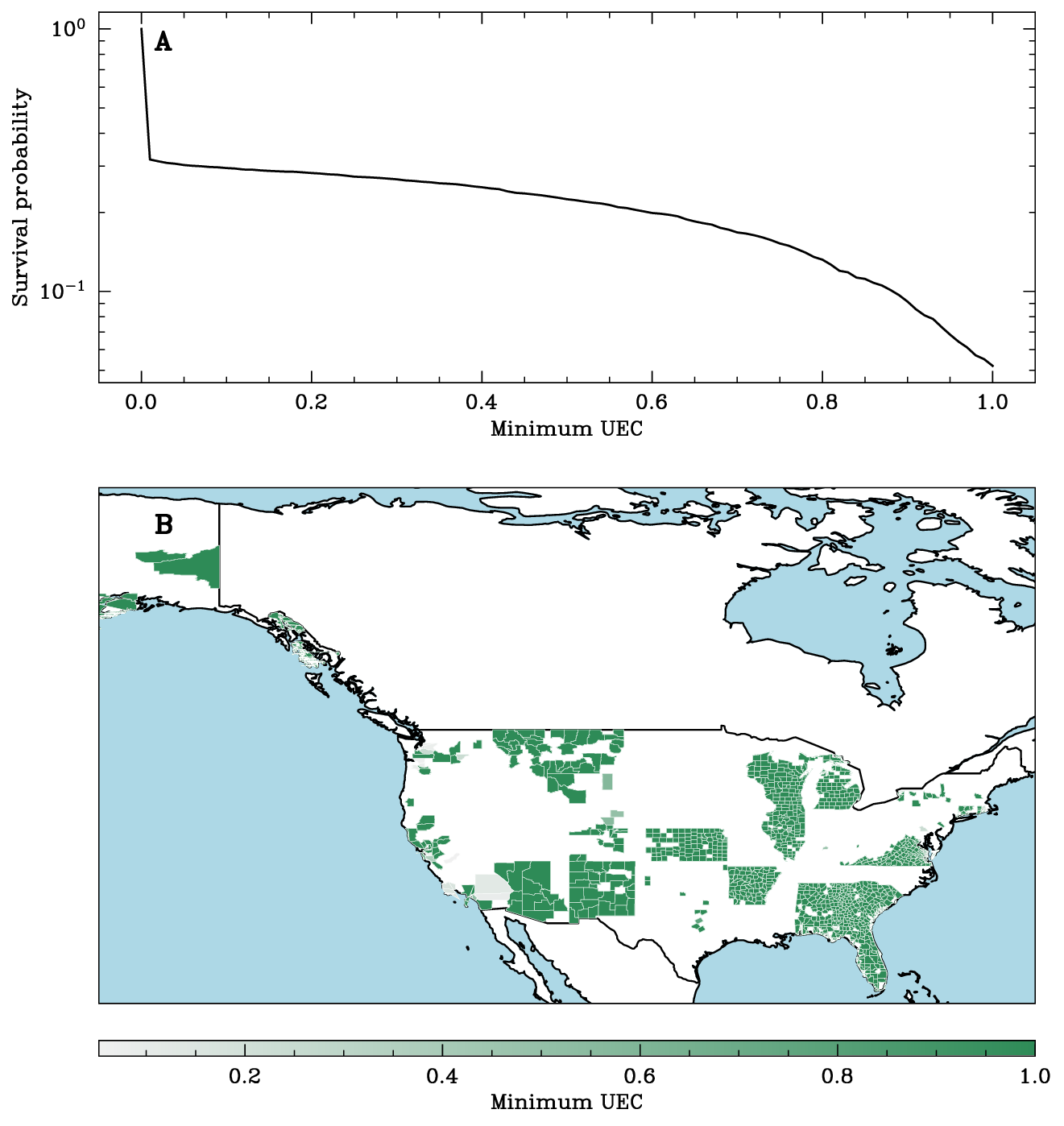}
    \caption{Visualization of each county's minimum UEC. For each county, we have computed the lowest UEC value between January 2018 and March 2023. This is 
    after sub-setting our data to nonfatal opioid occurrences but prior to filtering out counties with the threshold of $75\%$ minimum UEC. 
    \textbf{A)} Empirical survival function (i.e., complementary cumulative distribution) of the minimum UEC.
    \textbf{B)} Heatmap of the minimum UEC across U.S. counties. White indicates the absence of data (i.e., no coverage in any time period). The lightest grey-green indicates a minimum UEC of zero (i.e., no coverage in at least one but not all time periods).
    }
    \label{fig:uec_map}
\end{figure}

\clearpage
\subsection{Sensitivity to the UEC threshold\label{sec:sensitivity}}

In Section \ref{sec:ems}, we exclude counties with the minimum UEC below 75\%, following earlier work with this dataset \citep{casillas2022patient}. We note that, if $G_\beta$ is the overdose journey network arising from our dataset using a minimum UEC threshold $\beta$, then, for $\beta_1 > \beta_2$, $G_{\beta_1}$ is a subgraph of $G_{\beta_2}$. Therefore, we carry out the following three sensitivity analyses to confirm that our main results are not too strongly sensitive to our choice of the $\beta$ value (i.e., 75\%).

\begin{figure}[h!]
    \centering
    \includegraphics[width=1.0\textwidth]{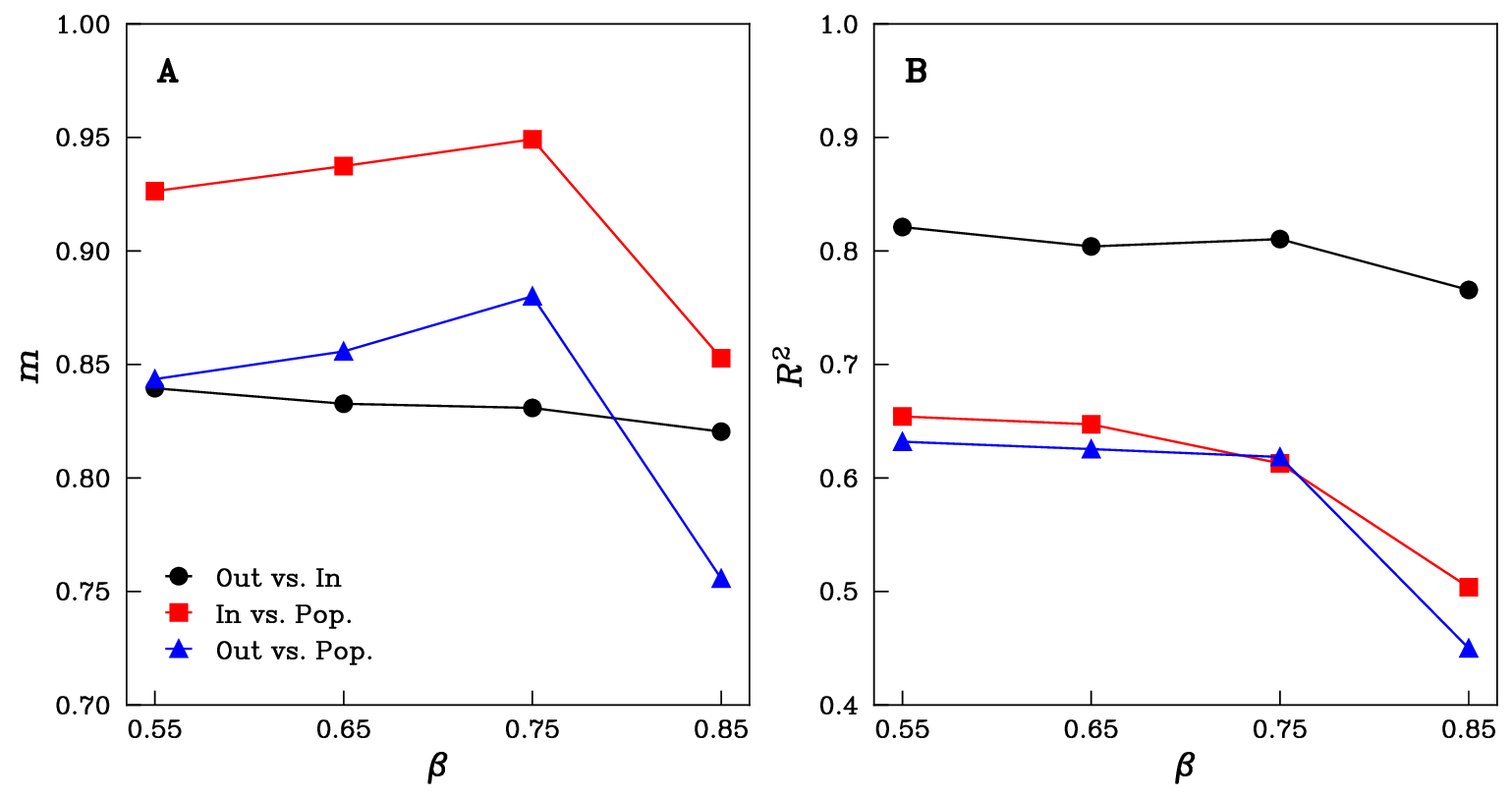}
    \caption{
    Dependence of the pairwise correlation between the weighted in-degree, weighted out-degree, and population nodes on the UEC threshold.
    \textbf{A)} Linear regression slopes ($m$) between the weighted out-degree and the weighted in-degree (Out vs. In), between the weighted in-degree and the population (In vs. Pop.), and between the weighted out-degree and the population (Out vs. Pop.) in the log-transformed space, as in Figure \ref{fig:degree_corr_weighted}.
    \textbf{B)} Correlation coefficients, $R^2$, for the same linear regression.
    Slopes and correlation coefficients involving the population are dampened with $\beta=0.85$. Otherwise, $m$ and $R^2$ values are generally consistent across values of $\beta$.
    }
    \label{fig:deg_sens}
\end{figure}

\begin{figure}[h!]
    \centering
    \includegraphics[width=1.0\textwidth]{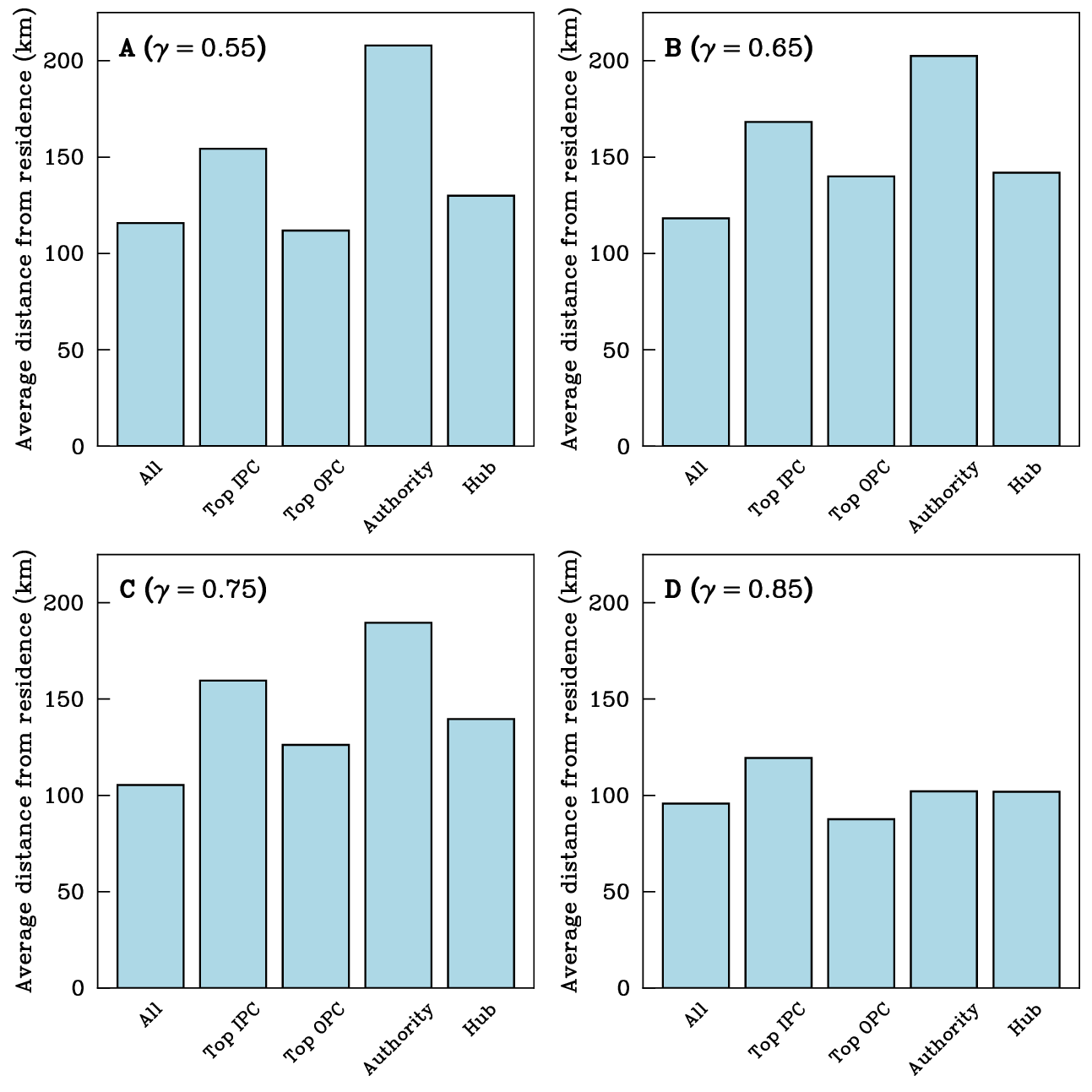}
    \caption{
    Dependence of the county's average overdose journey distance on the UEC threshold. \textbf{A)} $\beta=0.55$. \textbf{B)} $\beta = 0.65$. \textbf{C)} $\beta = 0.75$. \textbf{D)} $\beta = 0.85$. 
    Panel C replicates Figure \ref{fig:distances_static}D in the main text. 
    The results are similar among $\beta=0.55$, $0.65$, and $0.75$, whereas the results for the top authority counties are dampened for $\beta=0.85$.}
    \label{fig:dist_sens}
\end{figure}

\begin{figure}[h!]
    \centering
    \includegraphics[width=1.0\textwidth]{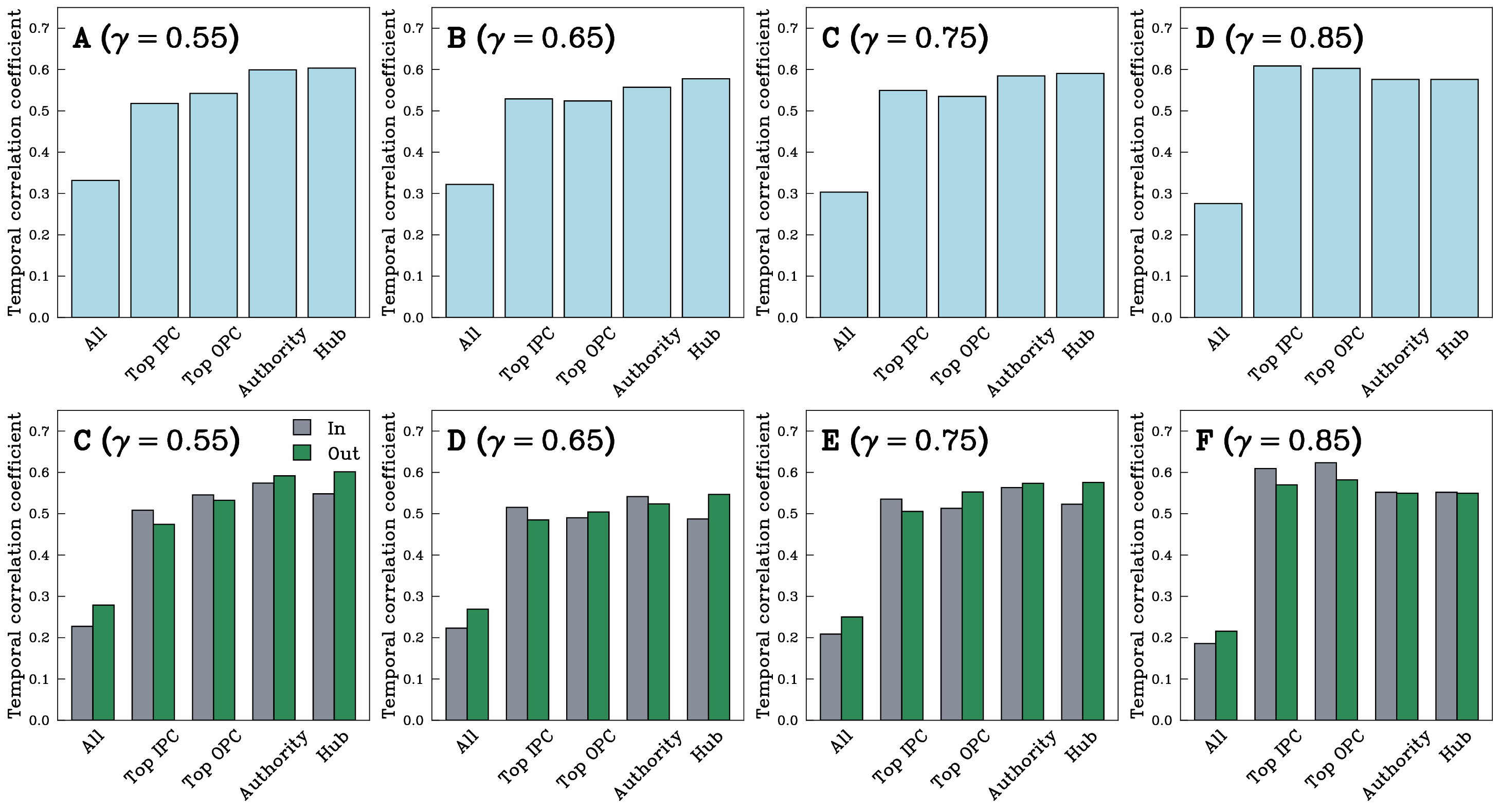}
    \caption{
    Dependence on the county's temporal correlation coefficient on the UEC threshold. \textbf{A)} Undirected, $\beta=0.55$. \textbf{B)} Undirected, $\beta=0.65$. \textbf{C)} Undirected, $\beta=0.75$. \textbf{D)} Undirected, $\beta=0.85$. \textbf{E)} Directed, $\beta=0.55$. \textbf{F)} Directed, $\beta=0.65$. \textbf{G)} Directed, $\beta=0.75$. \textbf{H)} Directed, $\beta=0.85$. 
    Panels C and E replicate Figures \ref{fig:t_corr}A and B, respectively, in the main text.
    We find that the results are qualitatively consistent across these values of $\beta$.}
    \label{fig:tcc_sens}
\end{figure}

First, we examine sensitivity of the relationships between the weighted in-degree, weighted out-degree, and population of counties (examined in Figure \ref{fig:degree_corr_weighted}) to the variation in the $\beta$ value. Figure \ref{fig:deg_sens}A and B shows that the slope and correlation coefficient, respectively, of the linear regression vary little, except those involving the population when $\beta = 0.85$ (red and blue lines in the figure). Second, our distance-to-overdose results (examined in Figure \ref{fig:distances_static}D) are qualitatively similar between $\beta=0.55$ and $\beta=0.75$ (see Figure~\ref{fig:dist_sens}A--C), whereas the prominence of the top authority counties is lost at $\beta=0.85$ (Figure~\ref{fig:dist_sens}D). Third, we analyze the edge persistence (examined in Figure \ref{fig:t_corr}) for different values of $\beta$ to find that the results are qualitatively comparable across the $\beta$ values (see Figure \ref{fig:tcc_sens}). Specifically, the top importers and exporters show notably greater temporal correlation coefficients than the network's average, in both undirected and directed settings.

With Figures~\ref{fig:deg_sens}, \ref{fig:dist_sens}, and \ref{fig:tcc_sens}, we conclude that the results are reasonably robust against variation in the $\beta$ value, at least between $\beta=0.55$ and $\beta=0.75$.

\clearpage
\subsection{Top \textit{k} selection\label{sec:kneedle}}

Throughout this work, we consider top $k$ counties across four metrics. To set the $k$ value, we first sort the weighted in-degree per capita, weighted out-degree per capita, hub score, and authority score, of all the counties. Then, for each of these metrics, we calculate the elbow point (i.e., the metric value yielding the maximum curvature) using the offline Kneedle algorithm. We set the algorithm's sensitivity parameter $S$ to $0$, which is recommended in perfect information settings \citep{satopaa2011finding}.

\begin{figure}[h!]
    \centering
    \includegraphics[width=1.0\textwidth]{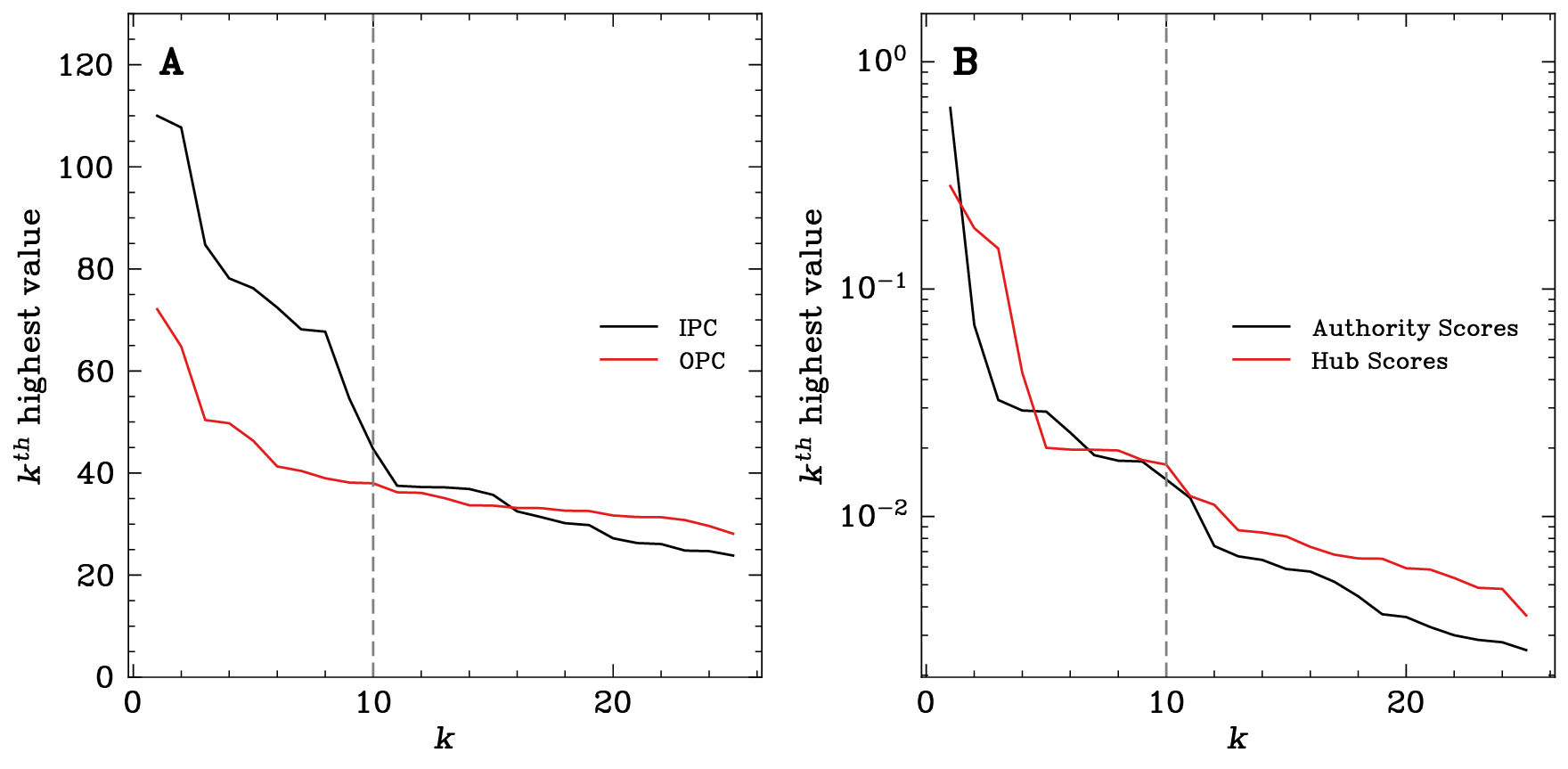}
    \caption{Sorted IPC, OPC, authority, and hub values. \textbf{A)} Sorted IPC and OPC values. \textbf{B)} Sorted authority and hub scores. 
    We selected $k=10$, indicated by the dashed lines, because it is a round number near the IPC and OPC elbow values.
    }
    \label{fig:elbows}
\end{figure}

This process reveals elbow values of 11, 10, 3, and 4 for weighted in-degree per capita, weighted out-degree per capita, authority scores, and hub scores, respectively (Figure \ref{fig:elbows}). In the interest of readability, interpretability, and consistency, we use $k=10$ for all the four metrics. This $k$ value is a round number close to or equal to the elbow values for the IPC and OPC, respectively. It is important to use the same $k$ value across the four measures because we compare observations from the top $k$ counties (e.g., average distance of geographically discordant overdose journey over the top $k$ counties) across the four measures.

\clearpage
\subsection{Unweighted degree analysis}

\begin{figure}[h!]
    \centering
    \includegraphics[width=1.0\textwidth]{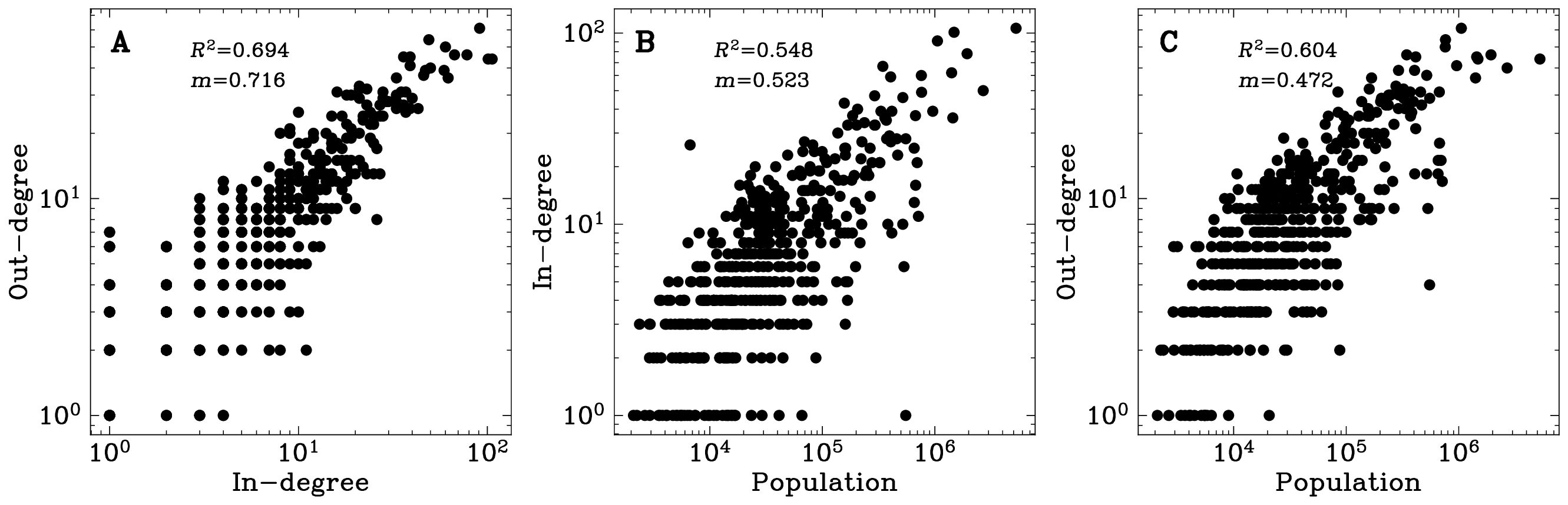}
    \caption{
    Relationships between the unweighted in-degree, unweighted out-degree, and population of the node.
    \textbf{A)} Relationships between the unweighted in-degree and the unweighted out-degree in the overdose journey network excluding self-loops. \textbf{B)} Relationship between the unweighted out-degree and the population. \textbf{C)} Relationships between the unweighted in-degree and the population. 
    The unweighted in- and out- degrees sub-linearly depend on the population size. The slopes of the linear regression in the log-transformed space are smaller than those for the weighted degree shown in Figure~\ref{fig:degree_corr_weighted}.
    In-degree and out-degree values have been incremented by one to ensure visibility in log-transformed space. 
    The $R^2$ values represent correlations between log-transformed values to limit distortion from outliers. The $m$ values represent the slopes of the linear regression in the log-transformed space.}
    \label{fig:degree_corr_unweighted}
\end{figure}

\clearpage
\subsection{County profiles with uncollapsed urbanization categories\label{sub:6-category-results}}

\begin{table}[h!]
    \centering
        \caption{Urbanization profiles of the top ten counties by in-degree per capita (top IPCs) and out-degree per capita (top OPCs), compared to the remaining counties. Values are not population-weighted.
        }
    \begin{tabular}{|l|l|>{\centering\arraybackslash}p{0.15\linewidth}|>{\centering\arraybackslash}p{0.15\linewidth}|>{\centering\arraybackslash}p{0.15\linewidth}|}
    \hline 
         & Category & \textbf{Top IPC} & \textbf{Top OPC} & \textbf{Other} \\
    \hline
        \textbf{Urbanicity} & Large central metro & 40.0 \% & 30.0 \% & 1.290 \% \\
        & Large fringe metro & 30.0 \% & 60.0 \% & 13.76 \% \\
        & Medium metro & 30.0 \% & 10.0 \% & 12.26 \% \\
        & Small metro & 0.0 \% & 0.0 \% & 12.26 \% \\
        & Micropolitan & 0.0 \% & 0.0 \% & 17.20 \% \\
        & Noncore & 0.0 \% & 0.0 \% & 43.23 \% \\
    \hline
    \end{tabular}
    \label{tab:degree_demographics_uncollapsed}
\end{table}

\begin{table}[h!]
    \centering
    \caption{Urbanization profiles of the top ten authorities and
    hubs, compared to the remaining counties. Values are not population-weighted.
    }
    \begin{tabular}{|l|l|>{\centering\arraybackslash}p{0.15\linewidth}|>{\centering\arraybackslash}p{0.15\linewidth}|>{\centering\arraybackslash}p{0.15\linewidth}|}
    \hline
         & Category & 
         \textbf{Authority} & \textbf{Hub} & \textbf{Other} \\
    \hline
    \textbf{Urbanicity} & Large central metro & 30.0 \% & 20.0 \% & 1.496 \% \\
        & Large fringe metro & 70.0 \% & 60.0 \% & 13.68 \% \\
        & Medium metro & 0.0 \% & 10.0 \% & 12.82 \% \\
        & Small metro & 0.0 \% & 10.0 \% & 11.97 \% \\
        & Micropolitan & 0.0 \% & 0.0 \% & 17.09 \% \\
        & Noncore & 0.0 \% & 0.0 \% & 42.95 \% \\
    \hline
    \end{tabular}
    \label{tab:hits_demographics_uncollapsed}
\end{table}

\clearpage
\subsection{Statistical results}\label{sec:stats}

\begin{table}[h!]
    \centering
    \begin{tabular}{|c|>{\centering\arraybackslash}p{0.15\linewidth}|>{\centering\arraybackslash}p{0.15\linewidth}|>{\centering\arraybackslash}p{0.15\linewidth}|>{\centering\arraybackslash}p{0.15\linewidth}|}
        \hline
         & \textbf{Population} & \textbf{Race \& Ethnicity} & \textbf{Employed} & \textbf{Poverty} \\
        \hline
        Top IPC vs.\,Top OPC &  & *** & * & *** \\
        Top IPC vs.\,Other & *** & *** & *** &  \\
        Top OPC vs.\,Other & *** & *** & *** & *** \\
        \hline
        Authority vs.\, Hub & *** & *** & *** & *** \\
        Authority vs.\,Other & *** & *** & *** & ** \\
        Hub vs.\,Other & *** & *** & *** & *** \\
        \hline
    \end{tabular}
    \caption{Statistical results for pairwise comparison between any two of the top IPC, top OPC, and the other counties and between any two of the top authority, top hub, and the other counties. 
    In terms of the socioeconomic variables evaluated, all pairwise comparisons from our HITS analysis and all but one pairwise comparisons from our IPC/OPC degree analysis indicate significant differences at the $\alpha=0.05$ significance level.
    We ran the Tukey's HSD test, which corrects for multiple comparisons \citep{tukey1949comparing, kramer1956extension}, for ``Population,'' which refers to the total population size. We used the contingency table-based G-test of independence \citep{mcdonald2014g} for ``Race \& Ethnicity,'', ``Employed,'' and ``Poverty.''
    *: $p<0.05$, **: $p<0.01$, and ***: $p<0.001$.
    All the results obtained with the G-test are Holm-Bonferroni-corrected for $3 \binom{3}{2}=9$ comparisons \citep{holm1979simple}.
    }
    \label{tab:sig_ipcsopcs_hits}
\end{table}

\begin{table}[h!]
    \centering
    \begin{tabular}{|c|>{\centering\arraybackslash}p{0.1\linewidth}|>{\centering\arraybackslash}p{0.1\linewidth}|>{\centering\arraybackslash}p{0.1\linewidth}|>{\centering\arraybackslash}p{0.1\linewidth}|>{\centering\arraybackslash}p{0.1\linewidth}|}
        \hline
        & \multicolumn{2}{c|}{\textbf{Journey Distance}} & \multicolumn{3}{c|}{\textbf{Temporal Correlation}} \\
        \hline
         & \textbf{Monte Carlo U-Test} & \textbf{Z-Test} & \textbf{HSD Test (Undirected)} & \textbf{HSD Test (In)} & \textbf{HSD Test (Out)}\\
        \hline
        All vs.\,Top IPC & *** & *** & ** & *** & **  \\
        All vs.\,Top OPC & *** & *** & * & *** & ** \\
        All vs.\,Authority & *** & *** & ** & *** & *** \\
        All vs.\,Hub & *** & *** & ** & *** & *** \\
        Top IPC vs.\,Top OPC & *** & *** & &  &  \\
        Top IPC vs.\,Authority &  & *** & & &  \\
        Top IPC vs.\,Hub &  & *** & & & \\
        Top OPC vs.\,Authority & *** & *** &  & & \\
        Top OPC vs.\,Hub & *** & ** & & & \\
        Authority vs.\,Hub &  & *** & & & \\
        \hline
    \end{tabular}
    \caption{Statistical results for pairwise comparisons of the journey distances and average temporal correlation coefficients.
    All pairwise comparisons indicate significant differences in the mean journey distance at the $\alpha=0.01$ significance level.
    The ``Monte Carlo U-Test" and ``Z-Test" columns reflect the results for journey distances, for which we ran the Monte Carlo U-test and Z-test described in Section \ref{sec:length_meth}. The ``HSD Test (Undirected)," ``HSD Test (In)," and ``HSD Test (Out)" columns reflect results for undirected temporal correlation coefficients, in- temporal correlation coefficients, and out- temporal correlation coefficients, respectively, for which we ran Tukey's HSD test \citep{tukey1949comparing, kramer1956extension}.
    *: $p<0.05$, **: $p<0.01$, and ***: $p<0.001$. 
    All the results are Holm-Bonferroni-corrected for $\binom{5}{2}=10$ comparisons \citep{holm1979simple}, with the exception of HSD tests, which already correct for multiple comparisons.
    }
    \label{tab:sig_dist_corr}
\end{table}

\end{appendices}

\clearpage
\bibliography{references}

\end{document}